# Revisiting the notion of "resonance wood" choice: a de-compartementalised approach from violin makers' opinion and perception to characterization of material properties' variability

C. Carlier[1], A. Alkadri[1], J. Gril[1,2] and I. Brémaud[1]

[1]Wood Team, Laboratory of Mechanics and Civil Engineering – LMGC, CNRS, UMR 5508, UM – CC048, Montpellier, France

[2]Institut Pascal, CNRS, UMR 6602, Univ. Clermont Auvergne, France

**Abstract:** This work aims to improve our understanding of the resonance wood and to investigate the interactions between their physical-mechanical properties, natural variability, and the violin makers' ways of choosing their materials. In order to identify the luthiers' practices and opinions, a socio-technical survey was conducted. Physical, vibrational, and visual/structural characteristics of the resonance wood obtained from several provenances with a variety of quality grades were also determined. Finally, these two approaches were completed by a psychosensory evaluation to compare the measurements that we have conducted with the perception and qualification of wood by the violin makers.

## 1. Introduction

Woods, as the main constitutive materials of many musical instruments, play an essential role in defining their "identity". The choice of wood by instrument makers is, therefore, a central question. It is intricately linked with the flora availability (local or imported), with cultural aspects both of woodworking craftsmanship and of aesthetics (musical and visual), as well as with physical behaviour and the acoustic response resulting from the material-structure-excitation system. This means that the study of instrument-making wood is intrinsically linked to the notion of diversity, may it be biological, cultural, functional, physical diversity or all these together. It also means that various different disciplines, often very distantly related to each other, may be involved in the study of wood choice by instrument-makers. However, most existing research usually adopt a single, or a limited number of viewpoints.

In some of our previous works, we aimed at relating biological and cultural diversity together with material physical/acoustical properties, in order to understand wood choice through different geocultural areas (for tuned idiophones' bodies or for chordophones' soundboards [1]), or through centuries (for European bows [2]). However, when one tries to consider jointly all the above-cited types of diversity together, there are often some missing information, at least from one of the involved disciplines or viewpoints. Especially, it seems that extra-European instrument-making cultures have benefited from less academic research than Western ones did, and early-music (or traditional) instrument-making less than Classical or modern ones. Direct interaction with instrument-makers from the various cultures is also more difficult (or even impossible for past cultures). Similarly, tropical, Mediterranean, or even temperate but "secondary" forest species have benefited from much less physical characterisation of their woods than did more common temperate hardwoods and especially softwoods. Therefore, in order to better understand the concept of "choice of wood by instrument-makers", with enough precision and information within and between the different fields involved, it may be useful to focus and to reduce the object of study while keeping the variety of viewpoints [3]. It is then necessary to select a case study which is well enough documented in different disciplines, in order to serve as a model for a decompartementalised approach.

Among all the many instrument-making woods or tonewoods around the World, the term "resonance woods" usually refers to those employed in the making of the violin family [4-7]. The choice is nearly exclusive to spruce (usually *Picea abies*) for the soundboard and to maple (usually *Acer pseudoplatanus*) for the back and sides. Due to the cultural importance and the "prestige" of the violin in Western classical/academic music, the instruments themselves, and the woods used for making them, have benefited from a lot of researches, both in organology/musicology, acoustics, forestry/wood science, much more than any other type of instrument-wood association. As an example, the sole *Picea* genus (spruce) is described by a number of tests for vibrational properties that is as big as the total number of such tests for all species (several tens of thousands) from tropical forests… [8]. And yet, the actual span and the different scales of variability within spruce resonance wood are still not completely assessed. There exist also several technical treatises of violin-making from the two past centuries which describe some criteria for choosing wood [7, 9]. Most importantly, violin-making encompasses a wide and vivid community of active luthiers, and, so it seems, a quite strong transmission of craftsmanship knowledge with little discontinuities could be observed. Therefore, the violin makers' choice for resonance wood would give us a good model to study the different facets involved in the choice of wood for musical instrument making. However, most researches have remained quite compartmentalized, with little integration between information from different disciplines, and even less so between scientific and craftsmanship knowledge. Besides that, a possible pitfall to watch for within this subject is that the prestige of the violin has generated a certain number of myths conveyed to the general public [10]. This is outside the topic of our work, but it suggests to conduct the study on resonance wood selection with a pair of fresh eyes.

The general objective of this research is therefore to revisit the concept of "resonance wood choice" by taking jointly into account the opinion and practice of violin-makers, the characterisation of physical-mechanical properties according to several scales of wood variability, and the sensory perception by makers in order to determine the relationships between the craftsmanship experience and the quantifiable properties of wood.

## 2. A survey to grasp luthiers' opinions, practices, and craftsmanship knowledge on wood

Although some aspects of the wood choice by violin-makers are written in early technical treatises and reviews [7, 9], there are very few formal research conducted recently to grasp the opinion of contemporary instrument makers. One of the rare example of a recent survey with violin makers, in the field of anthropology of techniques, examines the learning process and the transmission of the knowledge of the trade [11], but it seems that the knowledge of the wood itself was not specifically addressed in such surveys.

To improve our knowledge of violin-makers practices and of their main issues and opinions about wood, we created a specific survey [9, 12, 13]. The questionnaire was first designed for face-to-face interviews, following a modular structure, then was put online. It contains nine sections organized as follows:
   a. profile of the interviewee
   b. concept of the 'quality' of an instrument
   c. supply & wood choice in general
   d. evaluation of wood for top plates
   e. evaluation of wood for back plates
   f. evaluation of wood for bows
   g. aging, treatments and varnishes
   h. link to scientific and historical research
   i. questions, feedbacks and comments.

The objective is to capture how the luthiers consider the role of wood in the quality of violin and their procedure for obtaining wood supplies, as well as to categorize the different criteria they rely on for the selection of wood blanks.

The analysis of the survey is currently based on 15 complete responses of makers of the bowed-strings family. All of the participating craftsmen are French and produce string quartet instruments (violin, viola and cello, with less than 1/4 working on double-bass). Two of them are also luthier-bow makers, and another two also build early-music instruments (*violas da gamba* or early bows). Almost all of them are independent craftsmen. The majority (64%) of the makers work alone in their workshop. The average age of respondents is 46 years. Thus, the panel's experience repartition between the classes (less than 4 to 37 years) is balanced, with a majority having more than 15 years' experience. The panel of respondent luthiers are primarily engaged in the construction of new instruments (60% of their time in average).

The approaches that luthiers reckon to use during their work are "principally" or "often" (85%) empirical (know-how and technical knowledge) followed by historical (70% "often") but scientific approaches are only "moderately" or "little" (79%) used in daily work. Meanwhile, most of these makers reported a lot of interest for scientific research on instruments (76%) and/or on tonewood (83%). Individual makers show various "profiles" of interest between several fields of academic research (humanities and arts, natural or physical sciences). And many of them (70%) consider that they conduct research work in their workshop.

According to the interviewed luthiers, the choice of wood for top and back plates ranks within the top most important making parameters that determine the quality of an instrument. For "overall quality", they consider the wood for back plate to rank 1$^{st}$ (100% "very important" or "important") followed by varnish and wood for top plate (93% each). For "sound quality" also the choice of wood for the back comes 1$^{st}$, ex-aequo with fine adjustments (100% "important/very important") and the wood for top plate 2$^{nd}$ (93%), that is, even before important structural making parameters such as design, geometry and pre-stresses (86% each).

Most of the makers (80%) buy the wood for top and back-plates from specialized suppliers of tonewood. They generally acknowledge their suppliers' expertise and thus trust in their work and preselection of wood. While the luthiers are interested in the forest dimension, they do not consider themselves competent enough and/or do not have time to delve more deeply into it. It was confirmed by an interview with a supplier, who stated that makers are not interested enough to go to the forest. Some of the makers have also evoked the image of luthiers prospecting wood directly in forest as a myth, but one that they take pleasure to maintain.

The two most important criteria in the purchase of wood supplies by luthiers are the quality followed by price. The provenance and the drying time of the offered stocks have only secondary impact on the choice of suppliers, but the traceability of their stock has also sometimes emerged spontaneously. Moreover, suppliers also evoke the increase of the expectancy level of the makers. Previously, suppliers only proposed two categories of wood blanks to the makers based on their quality (one priced at 30 €, the other at 50 €). However, nowadays they have to offer a wider range of products with prices ranging from 10 € to 300 €. For spruce wood used for top plates, three main provenances were mentioned by the makers: Italy ("Val di Fiemme", "Italian Alps", "Italian Dolomites"), Switzerland ("Tyrol", "Swiss Alps") and France ("Jura", "French Alps"). This is consistent with the spruce tonewood repartition area but does not cover all the known provenances of such wood. French makers mainly focus on the historical cradle of violin-making and on the West part of tonewood repartition area, the nearest to them. For maple, the stated degree of precision for provenance varies greatly

(from "indifferent" or "Europe" to specific countries), with "the Balkans" being the most cited.

When choosing wood from the stock of a given supplier, interviewed makers rely mostly on their experience, on a "feeling acquired by practice" for evaluating wood quality, and only very little so on "rules from training or books" nor on "measuring tools". To guide their selection of wood pieces, they mostly use their personal evaluation of visual features (83% "very important" + 17% "important") and of weight/density (75% "very important + 25% "important"), while hand testing of acoustic features like "tap-tone" is stated as "important" (42%) but less strongly relied on (only 42% "very important"). Their choice is very little based on measuring tools (only 17% "important"), nor is it on the quality grades attributed by the suppliers (only 17% "important").

Concerning the choice of individual wood blanks (wedge shaped boards) for top plates, the selection criteria for spruce are mainly based, by order of stated importance, on the cutting plan/orientation (importance score 9.7/10), density (8.8), growth ring uniformity (8.3) and width (8.0), percentage of latewood (8.3), and hardness (8.0). All other stated criteria, including "tap tone", get importance scores lower than 7.0/10. For maple wood for back plates, the cutting plan also appears to be the most important criteria (8.5/10), followed by growth-ring width (8.0) and density (7.6), medular rays (7.1), and then only by figure (6.9). The cutting plan evaluation for both top and back plates is considered as "very important" or "important" and it might suggest a good perception of the mechanical performance of wood by the makers, since a minor change of grain angle is enough to greatly affect vibrational properties [14]. Most of the luthier's panel admit that they did change over time, at least a bit, their criteria for choosing a wood blank.

## 3. Material properties and natural variability of resonance woods

### 3.1. Characterisation of the physical, mechanical, acoustical, and structural properties

A wide sampling of wood sold for violin plate's blanks (wedge shaped boards) were collected from different suppliers, provenances, and with varying quality grades. Regarding the spruce, 49 top plates were gathered from three different provenances (France, Italy, Switzerland) and four quality grades (from the lowest - D - to the highest – A - quality grade). The selection of the top plates' blanks were not meant to be representative of a supplier's stock but, instead, to represent the maximum variability that he could offer. Similarly for maple, 24 blanks for back plates and sides were procured from four quality grades established by the suppliers, aiming to take into account the highest variability. One of the main selection criteria was the variety of wavy grain figures.

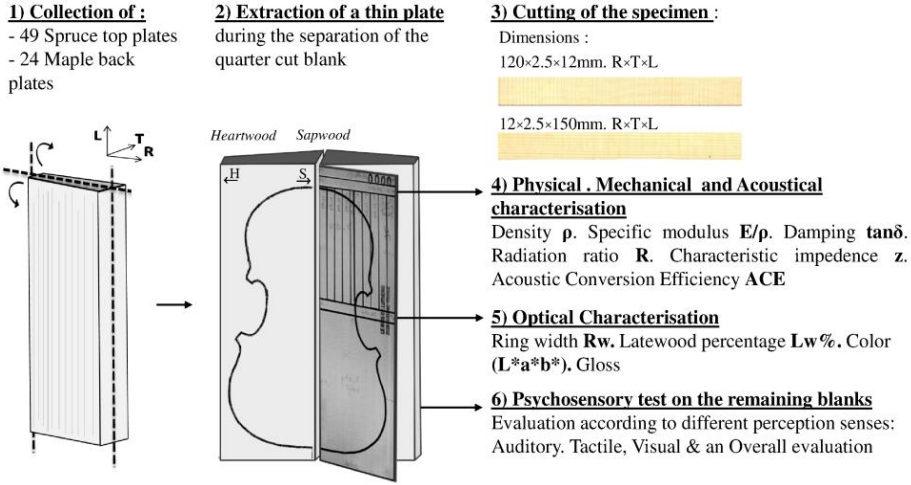

**Figure 1:** General sampling plan, specimen's cutting and test protocol

The sampling plan and specimen cutting protocol (Figure 1) aims at improving the knowledge of the multi-level variability of resonance wood properties, jointly with a study of makers' perceptive evaluation (see section 4). Compared to most other existing studies, here we also included an additional variability factor: the local variations of properties within a given wood blank. A thin board (2.5 mm in thickness) was extracted during the separation of the quarter cut blank wedge into two halves of the future top (soundboard) or back plate. These thin boards, representative of the initial violin-making blank wedges, were used for the laboratory testing while the remaining wedge blanks were kept for the psychosensory evaluation. Radial ($R$) and longitudinal ($L$) specimens were cut from the thin boards to assess the properties of the wood and their variations in those two directions. From each thin board, one to three radial specimens with the dimension of 120 mm × 2.5 mm × 12 mm (R × T × L) and seven to ten longitudinal specimens with the dimension of 12 mm × 2.5 mm × 150 mm (R × T × L) were obtained. They were conditioned at 20±1 °C and 65±2% RH for at least 3 weeks to reach the standard air-dry equilibrium moisture content (~12%). These specimens were characterized for their physical and vibrational properties (density $\rho$, specific modulus of elasticity $E'/\rho$, damping coefficient $\tan\delta$, in R and L direction, and also shear properties $G_{LT}$ and $\tan\delta_{GLT}$ on a sub-sampling), and their optical/structural traits (ring width $Rw$, latewood percentage $Lw$ %, colorimetry, gloss) as described in [9]. Then some acoustic "performance indexes" described in [1, 15-17] were calculated (radiation ratio $R=c/\rho$, characteristic impedance $z=\rho E$, acoustic conversion efficiency $ACE=R/\tan\delta$ and loudness index $L= R_L R_R/\delta_L \delta_R$).

**Table 1:** Summary of characterizations at the blank plates' level of spruce and maple resonance wood. Av: Average, S.Er: Standard error, Rov: Range of variation; COV: Coefficient of variation; Nb : Number of Top or Back plates; L: Longitudinal direction; R: radial direction; T Tangential direction

| | | Spruce | | | | | | | Maple | | | | | | |
|---|---|---|---|---|---|---|---|---|---|---|---|---|---|---|---|
| | | Min | Max | Av. | S. Er. | Rov (%) | Cov (%) | Nb | Min | Max | Av. | S. Er. | Rov (%) | Cov (%) | Nb |
| Mechanical vibrational properties | $\rho$ | 0.32 | 0.55 | 0.42 | 0.06 | 56 | 13 | 49 | 0.5 | 0.7 | 0.6 | 0.0 | 25 | 7 | 24 |
| | $E_L/\rho$ | 18.5 | 35.9 | 29.7 | 3.3 | 59 | 11 | 49 | 11.1 | 20.7 | 15.3 | 2.7 | 63 | 18 | 24 |
| | $E_R/\rho$ | 1.2 | 3.4 | 2.3 | 0.4 | 95 | 18 | 49 | 1.9 | 3.1 | 2.7 | 0.3 | 45 | 11 | 24 |
| | $\tan\delta_L$ | 0.006 | 0.010 | 0.007 | 0.001 | 53 | 11 | 49 | 0.01 | 0.01 | 0.01 | 0.00 | 53 | 16 | 24 |
| | $\tan\delta_R$ | 0.017 | 0.036 | 0.022 | 0.004 | 90 | 19 | 49 | 0.02 | 0.03 | 0.02 | 0.00 | 21 | 6 | 24 |
| | $E_L$ | 6.3 | 17.0 | 12.4 | 2.6 | 87 | 21 | 49 | 6.7 | 13.8 | 9.8 | 1.8 | 73 | 19 | 24 |
| | $E_R$ | 0.5 | 1.4 | 0.9 | 0.2 | 90 | 21 | 49 | 1.2 | 2.1 | 1.7 | 0.2 | 51 | 13 | 24 |
| | $G_{LR}$ | 0.70 | 1.22 | 0.93 | 0.16 | 55 | 17 | 9 | 1.2 | 1.6 | 1.4 | 0.1 | 27 | 8 | 8 |
| | $\tan\delta_{GLR}$ | 0.016 | 0.020 | 0.018 | 0.001 | 17 | 6 | 9 | 0.017 | 0.028 | 0.021 | 0.003 | 53 | 15 | 8 |
| F | $z_L$ | 1.5 | 3.1 | 2.3 | 0.4 | 71 | 17 | 49 | 2.2 | 4.5 | 3.4 | 0.7 | 68 | 20 | 24 |
| | $z_R$ | 0.4 | 0.8 | 0.6 | 0.1 | 64 | 15 | 49 | 0.8 | 1.7 | 1.3 | 0.3 | 68 | 22 | 24 |
| | $R_L$ | 9.7 | 16.6 | 13.3 | 1.5 | 52 | 12 | 49 | 4.8 | 7.6 | 6.1 | 0.7 | 45 | 12 | 24 |
| | $R_R$ | 2.3 | 5.5 | 3.7 | 0.6 | 88 | 17 | 49 | 2.1 | 3.2 | 2.6 | 0.2 | 40 | 10 | 24 |
| | $ACE_L$ | 1241 | 2628 | 1938 | 320 | 72 | 16 | 49 | 352 | 870 | 598 | 148 | 87 | 25 | 24 |
| | $ACE_R$ | 92 | 277 | 178 | 45 | 104 | 25 | 49 | 83 | 153 | 116 | 16 | 60 | 14 | 24 |
| | L | 16505 | 71047 | 35300 | 12356 | 155 | 35 | 49 | 3458 | 12498 | 7095 | 2460 | 127 | 35 | 24 |
| Optical features | Rw | 0.77 | 2.52 | 1.48 | 0.37 | 118 | 25 | 45 | - | - | - | - | - | - | - |
| | Lw % | 0.13 | 0.26 | 0.18 | 0.03 | 77 | 16 | 45 | - | - | - | - | - | - | - |

The summary of all measured properties of spruce and maple resonance wood (calculated by averaging values of the different specimens cut from a given plate) is provided in Table 1. The values of measured properties are consistent with other studies [4-6, 16-18], even if sometimes the ranges of variation that we observe are higher, resulting from our wide sampling.

In spruce we notably observe a large range of density $\rho$ (0.32 to 0.55), specific modulus of elasticity in the longitudinal direction $E'_L/\rho$ (18 to 35 GPa), longitudinal damping factor $\tan\delta_L$ (0.006 to 0.100), ring width $Rw$ (0.77 to 2.52 mm) and latewood percentage $Lw$ % (13 to 26 %). The expected properties for spruce, such as low damping coefficient ($\tan\delta$), low characteristic impedance ($z$), high axial specific modulus ($E'_L/\rho$) and high radiation ratio ($R$) were considered together with the role of the top-plate soundboard in the instrument (for instance, the higher the radiation ratio is, the greater the vibration amplitude and radiation will be).

Regarding maple wood for back plate, its range of variation (*Rov*) of the properties appears to be lower than the spruce's, which is probably due to, partly at least, the lower number of boards studied, for the coefficients of variation (*Cov*) are of comparable amplitude for the 2 species. As expected, maple specimens possess higher density, damping and characteristic impedance, while having lower specific modulus, radiation ratio and ACE than spruce. The measured properties of spruce and maple present specificities adapted to their role in the instrument making [15, 17, 19]. Some current studies have taken into account these differences in wood properties in the modelisation of violins in order to quantify the respective impact of various wood parameters on the vibroacoustic response of the instrument [20].

## 3.2. Natural variability of properties between different violin-making wood blanks

In this part, we will focus on spruce since the variability of maple properties was previously analysed in regards with its anatomical features in our recent paper [21]. Our general objective is to evaluate the variability of physical and acoustical properties of tonewood in relation with qualities for use. Figure 2 shows, for a given property, the differences between quality grades and whether or not they are significant using Multiple Comparison test. We could see that the differences between the first and second quality grades were not significant; significant differences were found on the fourth quality grade, which clearly differs from other grades. However, even without statistically significant differences between suppliers, clear trends can be observed among the four quality grades: with the increase in the quality grade, we would found lower density ($\rho$), higher specific modulus ($E'_L/\rho$), lower damping ($\tan\delta_L$), and lower latewood percentage (*Lw*, %) and ring width (*Rw*) within the resonance wood. Those observations are consistent with current makers' selection criteria and previous research, which however included fewer quality grades [18].

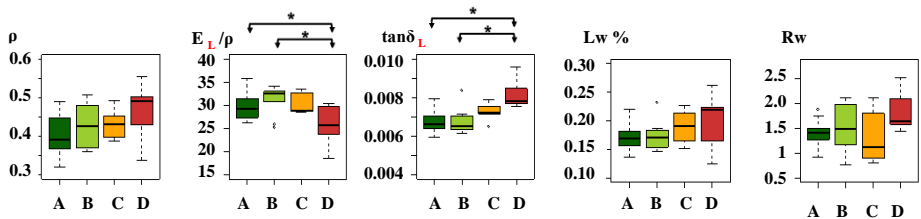

**Figure 2:** Distribution of some physical-mechanical properties and optical features of resonance spruce plates according to different quality grades (A=highest, D=lowest): the significance of differences between grades is indicated by a star (*).

It is also interesting to put those results in regards with the survey: although makers claimed their trust into their supplier's expertise, they also said that they paid little attention to their grades. We could explain this paradox by presenting this assumption: the suppliers sort resonance woods mostly on cutting, optical, and density criteria, which seemed to also lead to a mechanical and acoustical sorting. However, the distribution of the properties within each grade shows that it is indeed possible to find adequate resonance wood in other grades than the first one. Considering the fact that the quality grades also affect the prices, one can propose a hypothesis that the makers' behaviour on quality grades attributed by suppliers, is mostly guided by economic reasons.

To complete our understanding of the variability between different wood blanks, the correlation between optical-structural characteristics (latewood percentage and ring width) and physical-mechanical properties or acoustical indexes is represented according to wood provenance and to quality grade in figure 3. Latewood percentage highly correlates with wood density ($\rho$) and acoustical indexes such as radiation ratio ($R$) and impedance ($z$) in both longitudinal and radial directions for both the 3 provenances and the 4 quality grades. Concerning ring width ($Rw$), its correlations with other properties is weaker; however, ring width seems to be a good indicator of specific modulus ($E'_L/\rho$) in longitudinal direction. This last result, obtained from spruce specimens that have been pre-selected as resonance wood, is interesting because such a correlation barely exist in general supply (non selected wood) or faster grown trees of the same species [22].

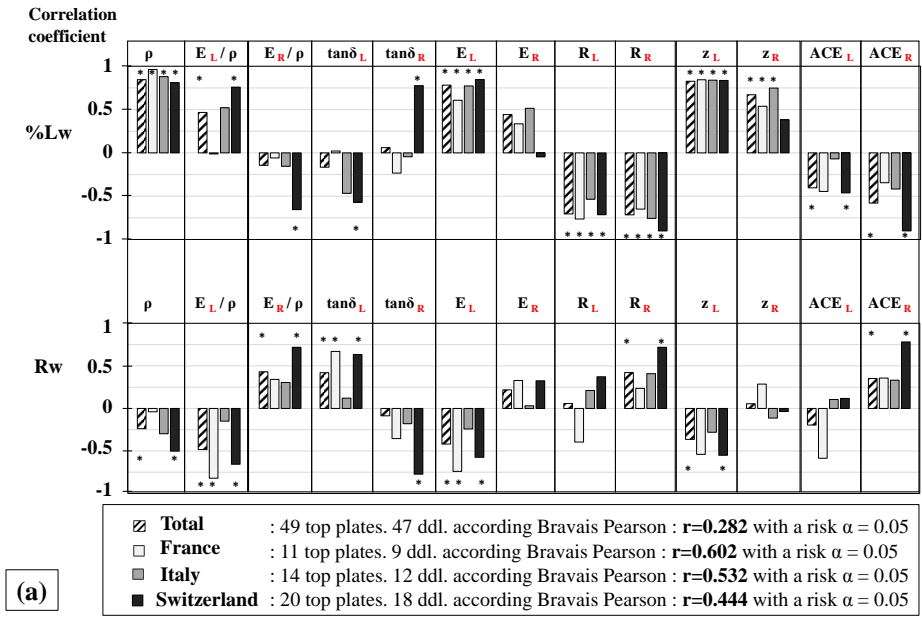

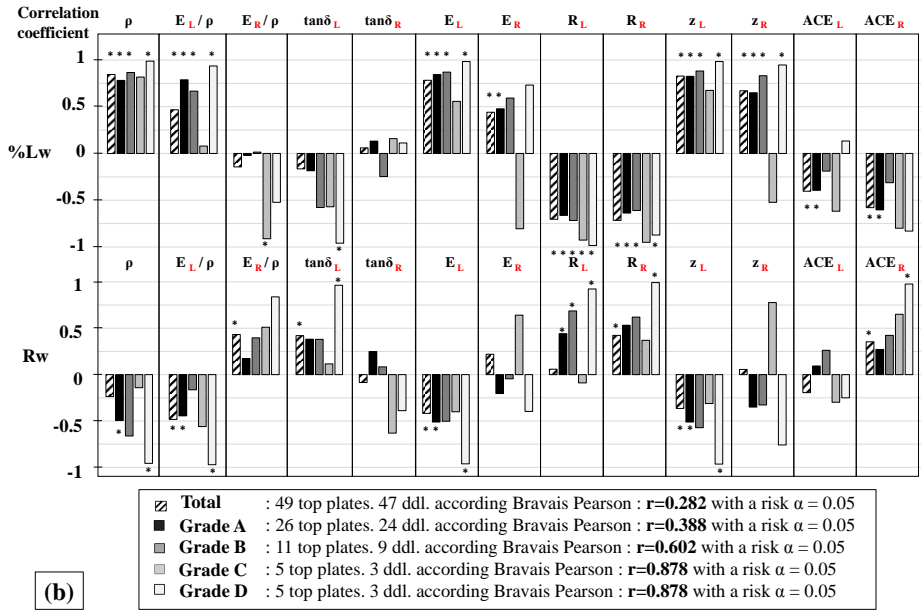

**Figure 2:** Correlations between visual features and mechanical and acoustical properties according to top plates provenance (a) and quality grade (b). The significance of correlations is indicated by a star (*).

However, there are clear differences in these correlations depending on the provenance. For example, while it seems that the ring width of the wood samples from Italy do not share any relationships with any properties, for wood samples from France and Switzerland, the ring width correlates with longitudinal specific modulus, damping, and anisotropic ratios. Further, the relationships between latewood and specific modulus also differ according to each provenance. As an example, the correlation is highly significant for top plates from Switzerland but non-existent for those from France.

To sum it up, instrument makers' criteria for visual evaluation seem to give us an idea of some of the physical/acoustical properties of wood, even if those relations are not always equal for every provenances and quality grades of the wood.

### 3.3. Variations within individual blank wedges of resonance spruce

Within-plate variations of wood properties, corresponding to different radial positions, are shown in Table 2. Although the mean level of variability (ROV or COV) is smaller within one plate than between different plates (Table 1), the maximum coefficient of variations behaves differently. Within-plate variation is equivalent to between-plate variation for longitudinal specific modulus (11% vs. 11 % between plates) and even significantly higher for latewood (28% vs. 16%) and ring width (39% vs 25%). For some properties, a higher variability can be observed within some given plates than between the different plates. These re-sults suggest that, in resonance wood study, intra-plate variations must be considered almost at the same level as the inter-plate variations, even if the amplitude depends on the studied plate.

Table 2: Variation within one spruce top plate of wood properties and optical features. Rov: Range of variation = (max-min)/average; Cov: Coefficient of variation = standard deviation/average.

|  |  | Rov | Cov |
|---|---|---|---|
| ρ | Average | 11% | 4% |
|  | Min | 3% | 1% |
|  | Max | 24% | 7% |
| $E_L/\rho$ | Average | 13% | 4% |
|  | Min | 4% | 2% |
|  | Max | 34% | 11% |
| $\tan\delta_L$ | Average | 14% | 5% |
|  | Min | 4% | 1% |
|  | Max | 26% | 11% |
| Rw | Average | 63% | 20% |
|  | Min | 27% | 8% |
|  | Max | 110% | 39% |
| Lw % | Average | 32% | 10% |
|  | Min | 11% | 4% |
|  | Max | 96% | 28% |

Figure 4 gives the radial variation of the main properties and characteristics: density ($\rho$), specific modulus ($E'_L/\rho$), damping ($\tan\delta_L$), ring width (Rw), and latewood percentage (Lw, %). They are shown from outerwood (closer to sapwood) to innerwood (towards center of the tree), corresponding to the center and the side, respectively, of the top-plate of a musical instrument such as a violin. From the center to the side (of a future soundboard), different trends are observed according to the property: generally an increase of specific modulus and ring width, and a decrease of density, damping and latewood percentage. However the existence of marqued heterogeneity, both concerning the profile and the amplitude of variations, can be noted depending on the wood plate.

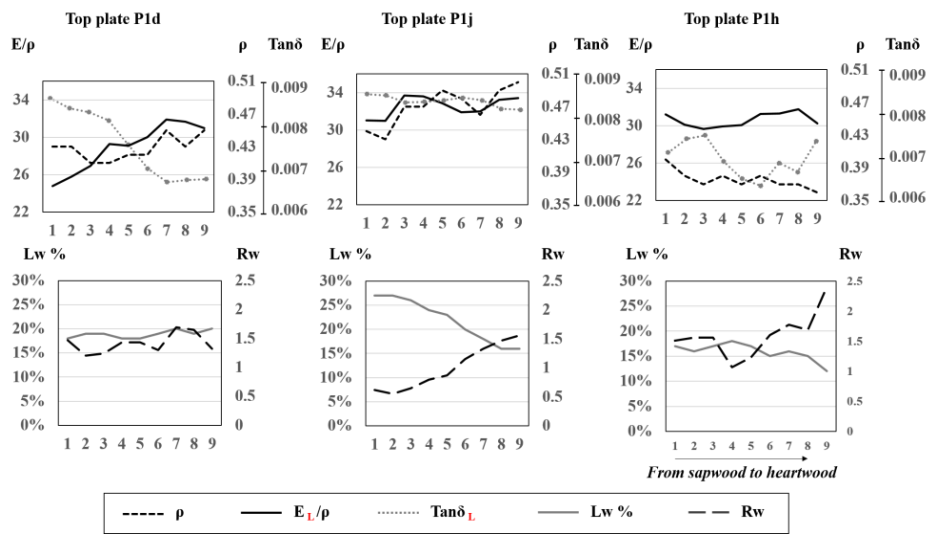

**Figure 3:** Representation of the intra-plate variability of some wood properties and optical features. The top plates presented here (P1d, P1j, P1h) belong to top quality grade A.

During the survey, some makers indicated a preference for resonance spruce with thinner rings in the center of the instrument (0.8 to 1.5 mm) and wider rings in the sides (2.0 to 2.5 mm). From our previous observation, one can wonder: what is the impact of the selection criteria based on ring width on the material properties? The intra-plate variability has to be taken into account since the overall characterisation of the board does not provide enough information on wood properties and their variations, and one unique sample from the board may not be representative of the top plate properties.

## 4. Sensory perception of resonance spruce

In the previous sections, we studied the tonewood selection from several viewpoints, but there was no direct confrontation between the makers, the material, and its measured variability. In this part of our work, we aim to understand the direct interactions between the craftsmen and their wood through a psycho-sensory evaluation. There are only few studies on the sensory perception of wood by instrument makers (e.g. [23]), with, up to our knowledge, a single one focusing on violin-making resonance wood [6], while it is suggested that trade-speciality strongly impact the wood perception [24].

As the perceptual criteria used for wood evaluation can be primarily based on the tactile, visual, and auditory sensory modes, we thus designed the test in four parts. The three first parts of the test were created to evaluate, separately, the respective contributions of the different perceptual modes. The final part of the test was an overall evaluation, aiming to reproduce, as closely as possible, the habits of luthiers in their workshop. Except for the last part, each protocol follows the same outline. First, some attributes are evaluated to establish a sensory profile of the wood blank according to each sense. This part allows us to characterize the product and to learn how the attributes may define a board's quality as excellent or unusable. The blanks were then evaluated by a general quality rating for each sense. Finally, an acceptance test was performed to determine the emotional attachment to the product.

In the first part of the protocol, the resonance woods were evaluated through hearing only. Makers were asked to evaluate several attributes such as 'pitch', 'sound duration', and 'crystallinity'. In the second part, the sense of touch was tested. The makers were asked to evaluate the top or back plates by touching the surface of the material only, then by weighing it. Makers had to evaluate 'roughness', 'hardness', and 'density' of the material. The third part focused on the vision. The panel observed the samples without touching it. The selected attributes were, for spruce top plates, 'ring width' and 'regularity', 'latewood percentage', 'cutting plan', 'colour', 'gloss', and 'hazelgrowth' (indented rings).

Finally, the overall evaluation was conducted by letting the makers use all their senses, as they would usually do in their workshop. This overall evauqlation included a quality rating, an acceptance test, and optionally a free verbalisation.

The relative weight of the 3 considered perception senses—used to examine the material's quality—and the attributes favourably perceived by different senses to define a good top plate were determined. The evaluation of the makers was analysed in regard to the physical measurements of the specimen in order to characterize their perception.

As the test was time-consuming (ranging from 40 minutes to two hours), the analyses of the psychosensory study on spruce are currently based on seven complete responses only. Therefore, the number of participants is not yet significant.

While observing the evaluation of each wood blanks according to the general quality rating for the different perception senses, we observed different kinds of variability interacting with the obtained results. Firstly, the perceived differences between plates differed according to each sense of evaluation: discrimination between different spruce blanks, for instance, was more difficult to access by the auditory rating than through the visual, tactile, or overall ratings (figure 5). This variability could be attributed to two causes: (1) the makers could indeed perceive more differences through the tactile, visual, or overall senses than by using their auditory senses only (which means that the auditory perception might have less weight than the tactile or visual perception); (2) the perceived differences could be due to a flaw in the conception of the protocol (the makers could have felt more confident during the successive steps of the test and thus give more and more adamant evaluation).

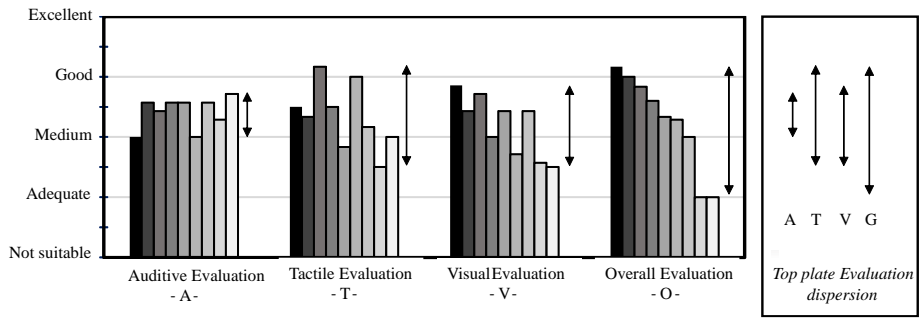

**Figure 4:** Top plates' average ratings and their variation range obtained during the four phases of the psychosensory test

Second, we observed a variability between the makers' behaviour. We observed diverse quality rating values according to the makers—some makers always gave lower ratings than others—and also differences in the distribution behaviour—for example, some makers tend to give wider differences in ratings between the various top plates.

**Table 3:** Correlations' table between sensory attributes and their quality rating evaluation. A value in bold indicates a significant correlation according to Pearson-Bravais table with an α-risk of 0,05

| Perceived attributes | | Correlation to : | |
|---|---|---|---|
| Perceived Pitch | Pitch | -0.04 | Auditory rating |
| Perceived Length | Len. | **-0.56** | |
| Perceived Crystallinity | Cry. | **0.47** | |
| Perceived Roughness | Rou. | -0.10 | Tactile rating |
| Perceived Hardness | Hard. | -0.30 | |
| Perceived Density | $\rho$ | **0.73** | |
| Perceived Ring width | Rw | **-0.46** | Visual rating |
| Perceived Regularity | Reg | -0.31 | |
| Perceived Cutting | Cut. | **-0.41** | |
| Perceived Latewood % | Lw% | 0.27 | |
| Perceived Gloss | Gloss | **-0.41** | |
| Perceived Color | Color | -0.20 | |
| Perceived Hazelgrowth | Hg. | -0.20 | |

To define a "good top plate", we first isolated the attributes that are favourably perceived by the makers by looking at the correation between individual attributes and the general quality rating for each sense. Afterwards, we analysed the perceived attributes in relation to the physical measurements that we had previously conducted on the specimen. The results (Table 3) show that the auditory quality rating of the wood blanks is mainly correlated to attributes that express the damping ("length/duration of sound" and "crystallinity"), which is consistent with results from the literature [23]. At the same time, according to the correlation with the measured wood properties (figure 6), it also appears that the auditory attribute which the makers perceived most is the "pitch", which is an indication of the resonance frequency or sound wave celerity. For a constant geometry—which is the condition of the material during our test—the higher the "pitch" of the fundamental note, the higher the speed of sound. In short, it is an indicator of the specific modulus of elasticity of the wood. However, on the daily basis, in the real conditions of workshops, the dimensions of the blank wedges for plates are highly varied. Therefore, this parameter (the pitch) cannot be taken into account in practice—unlike the damping, which is more independent of the geometry. And so the behaviour of the makers—not taking the pitch into account—is adapted to their usual work situation.

|  |  | Perceived | | | | | | | | | | | |
|---|---|---|---|---|---|---|---|---|---|---|---|---|---|
|  |  | Pitch | Len. | Cry. | Rou. | Hard. | ρ | Rw | Reg. | Cut. | Lw % | Gloss | Color | Hg |
| Optical characteristics measurement | Ring width | -0.57 | 0.6 | 0.33 | -0.55 | -0.43 | 0.33 | 0.93 | 0.6 | 0.59 | 0.23 | 0.38 | -0.28 | 0.09 |
|  | Latewood | 0.63 | -0.07 | -0.06 | 0.83 | 0.93 | -0.82 | -0.4 | -0.1 | 0.13 | -0.58 | -0.13 | 0.58 | -0.09 |
|  | ring width regularity | -0.33 | 0.79 | 0.37 | -0.25 | -0.2 | 0.11 | 0.78 | 0.47 | 0.55 | -0.43 | 0.67 | 0.25 | 0.23 |
|  | a* color parameter | 0.06 | -0.01 | -0.37 | 0.73 | 0.74 | -0.6 | -0.27 | -0.01 | 0.13 | -0.64 | 0.18 | 0.82 | 0.51 |
|  | b* color parameter | 0.24 | -0.02 | -0.2 | 0.78 | 0.82 | -0.66 | -0.33 | -0.14 | 0.04 | -0.68 | 0.09 | 0.71 | 0.34 |
|  | L* color parameter | 0.04 | -0.14 | 0.19 | -0.66 | -0.68 | 0.61 | 0.11 | 0.07 | -0.09 | 0.52 | -0.34 | -0.78 | -0.55 |
|  | Gloss | 0.9 | -0.1 | 0.23 | 0.59 | 0.44 | -0.53 | -0.55 | -0.38 | -0.12 | 0 | -0.63 | 0 | -0.02 |
| Longitudinal properties measured | Density ρ | 0.56 | -0.07 | -0.12 | 0.86 | 0.97 | -0.89 | -0.37 | -0.05 | 0.19 | -0.55 | -0.12 | 0.65 | 0.03 |
|  | $E_L/\rho$ | 0.92 | -0.43 | 0.21 | 0.43 | 0.41 | -0.49 | -0.67 | -0.56 | -0.37 | 0.15 | -0.8 | -0.25 | -0.39 |
|  | $tan\delta_L$ | -0.98 | 0.33 | -0.3 | -0.4 | -0.38 | 0.58 | 0.59 | 0.53 | 0.24 | -0.05 | 0.79 | 0.16 | 0.26 |
|  | $E_L$ | 0.85 | -0.34 | 0.01 | 0.75 | 0.81 | -0.8 | -0.63 | -0.36 | -0.11 | -0.26 | -0.54 | 0.25 | -0.21 |
|  | $z_L$ | 0.77 | -0.24 | -0.02 | 0.82 | 0.89 | -0.86 | -0.55 | -0.26 | 0 | -0.36 | -0.4 | 0.4 | -0.13 |
|  | $R_L$ | -0.03 | -0.17 | 0.24 | -0.59 | -0.78 | 0.69 | -0.06 | -0.29 | -0.47 | 0.76 | -0.35 | -0.87 | -0.29 |
|  | $ACE_L$ | 0.68 | -0.42 | 0.32 | -0.15 | -0.27 | 0.06 | -0.5 | -0.58 | -0.5 | 0.49 | -0.85 | -0.68 | -0.37 |
| Radial properties measured | $E_R/\rho$ | -0.75 | -0.1 | -0.5 | -0.41 | -0.33 | 0.33 | 0.32 | 0.43 | 0.29 | 0.12 | 0.17 | 0.07 | 0.46 |
|  | $tan\delta_R$ | -0.11 | 0.05 | 0.13 | -0.12 | -0.35 | 0.62 | -0.13 | -0.21 | -0.5 | 0.33 | 0.17 | -0.46 | -0.44 |
|  | $E_R$ | -0.09 | -0.13 | -0.55 | 0.49 | 0.64 | -0.58 | -0.07 | 0.34 | 0.49 | -0.39 | -0.02 | 0.68 | 0.52 |
|  | $z_R$ | 0.3 | -0.09 | -0.34 | 0.77 | 0.9 | -0.85 | -0.25 | 0.16 | 0.4 | -0.53 | -0.1 | 0.75 | 0.33 |
|  | $R_R$ | -0.82 | -0.02 | -0.23 | -0.75 | -0.77 | 0.79 | 0.39 | 0.27 | -0.01 | 0.38 | 0.26 | -0.35 | 0.17 |
|  | $ACE_R$ | -0.63 | -0.03 | -0.3 | -0.55 | -0.43 | 0.28 | 0.43 | 0.41 | 0.36 | 0.09 | 0.1 | 0.02 | 0.45 |
| Anisotropic ratios measured | $E_L/\rho / E_R/\rho$ | 0.91 | -0.22 | 0.37 | 0.42 | 0.37 | -0.41 | -0.56 | -0.55 | -0.39 | -0.05 | -0.51 | -0.17 | -0.52 |
|  | $tan\delta_R / tan\delta_L$ | 0.43 | -0.2 | 0.24 | 0.07 | -0.15 | 0.3 | -0.49 | -0.51 | -0.65 | 0.31 | -0.29 | -0.51 | -0.57 |
|  | $E_L/E_R$ | 0.89 | -0.26 | 0.34 | 0.42 | 0.39 | -0.41 | -0.58 | -0.57 | -0.43 | -0.03 | -0.51 | -0.19 | -0.56 |
|  | $z_L/z_R$ | 0.87 | -0.26 | 0.35 | 0.44 | 0.42 | -0.43 | -0.57 | -0.56 | -0.42 | 0.02 | -0.51 | -0.2 | -0.56 |
|  | $R_L/R_R$ | 0.94 | -0.12 | 0.42 | 0.44 | 0.36 | -0.44 | -0.51 | -0.51 | -0.31 | -0.04 | -0.51 | -0.14 | -0.42 |
|  | $ACE_L/ACE_R$ | 0.65 | -0.21 | 0.3 | 0.19 | 0.01 | 0.08 | -0.56 | -0.56 | -0.6 | 0.18 | -0.39 | -0.41 | -0.58 |

**Figure 5:** Correlations matrix between attributes perceived by makers and material characteristics and properties. A value in a black cell indicates a significant correlation according to Pearson-Bravais table with an α-risk of 0,05

Concerning the evaluation of wood through tactile sense, the main attribute to define a "good" wood blank was its density (R=0.73 between evaluation of attribute –from "very dense" to "very light"- and quality rating of blank plates), which was also well perceived by the makers (R=-0.89 between evaluation of attribute and measured property). However, during our test, since the geometry of the wood blanks was constant, the density evaluation was not made under the same conditions as the makers usually do. Their rating of density is directly due to an approximation of the perceived mass. It is thus questionable if, in real life condition with the blank-plates of varying dimensions, the estimation would have been as strongly correlated to actual measurements as the results that we obtained.

Next, the visual quality rating of top plates was mainly based on the perceived "ring width" (attribute also highly correlated to optical/structural measurements), and also to the attributes of "cutting plane" and "gloss". The perception of color and ring width by the makers is fairly accurate when compared to the physical measurements. Furthermore, their perception of attributes of latewood percentage and color gives them indirect information on the longitudinal radiation ratio, while the gloss is correlated with both longitudinal specific modulus and damping.

|  |  | Auditory Rating | Tactile Rating | Visual Rating | Overall Rating |
|---|---|---|---|---|---|
| Optical characteristics measurement | Ring width | 0.52 | 0.17 | 0.2 | -0.23 |
|  | Latewood | 0 | -0.68 | -0.63 | -0.52 |
|  | ring width regularity | 0.18 | -0.03 | 0.23 | -0.65 |
|  | a* color parameter | -0.33 | -0.61 | -0.09 | -0.58 |
|  | b* color parameter | -0.18 | -0.65 | -0.3 | -0.58 |
|  | L* color parameter | 0.15 | 0.73 | 0.13 | 0.61 |
|  | Gloss | 0.3 | -0.38 | -0.77 | -0.1 |
| Longitudinal properties measured | Density ρ | 0.01 | -0.77 | -0.61 | -0.56 |
|  | $E_L/\rho$ | 0.28 | -0.29 | -0.78 | 0.24 |
|  | $\tan\delta_L$ | -0.27 | 0.36 | 0.74 | -0.11 |
|  | $E_L$ | 0.11 | -0.6 | -0.77 | -0.17 |
|  | $z_L$ | 0.1 | -0.69 | -0.75 | -0.33 |
|  | $R_L$ | 0.2 | 0.67 | 0.11 | 0.78 |
|  | $ACE_L$ | 0.23 | 0.23 | -0.37 | 0.63 |
| Radial properties measured | $E_R/\rho$ | -0.37 | 0.27 | 0.77 | 0.1 |
|  | $\tan\delta_R$ | 0.11 | 0.54 | -0.08 | 0.42 |
|  | $E_R$ | -0.28 | -0.52 | 0.07 | -0.48 |
|  | $z_R$ | -0.13 | -0.74 | -0.33 | -0.61 |
|  | $R_R$ | -0.25 | 0.67 | 0.82 | 0.43 |
|  | $ACE_R$ | -0.3 | 0.24 | 0.76 | 0.06 |
| Anisotopic properties measured | $E_L/\rho / E_R/\rho$ | 0.27 | -0.24 | -0.78 | 0.1 |
|  | $\tan\delta_R / \tan\delta_L$ | 0.16 | 0.35 | -0.41 | 0.5 |
|  | $E_L/E_R$ | 0.26 | -0.24 | -0.78 | 0.14 |
|  | $z_L/z_R$ | 0.32 | -0.27 | -0.82 | 0.14 |
|  | $R_L/R_R$ | 0.33 | -0.28 | -0.81 | 0.02 |
|  | $ACE_L/ACE_R$ | 0.18 | 0.19 | -0.55 | 0.39 |

**Figure 6:** Correlations matrix between quality ratings evaluations by makers and the measured material characteristics and properties. A value in a black cell indicates a significant correlation according to Pearson-Bravais table with a α-risk of 0,05

Finally, in Figure 7 we could see the correlations between physical, mechanical and vibrational properties of wood with the general quality ratings evaluation according to the four modalities of the test (auditory, tactile, visual and overall). As expected, the tactile evaluation of quality rating is significantly correlated to the measured wood density. It is also well related to both longitudinal and radial wood vibrational properties and indexes (such as impedance or radiation ratio) and to optical characteristics such as latewood percentage and color. The visual quality rating evaluation shows a significant correlation with Young's modulus and specific modulus of elasticity in both longitudinal and radial directions and is also well related to damping and characteristic impedance. The overall quality rating evaluation of spruce plates by the makers is significantly correlated with longitudinal radiation ratio, considered to be a good criterion to describe the resonance woods. It appears that the makers' perception of visual and tactile attributes gives an indirect—but accurate—indication of the wood's physical and mechanical properties while the overall evaluation is a reliable indicator of the radiation ratio.

## Conclusion

The originality of our approach permitted us to not only explore the question of wood selection under different research areas that are complementary but yet usually compartmentalised, but also to bring a real qualified exchange according to the different points of view. The survey gave information on the criteria used by the luthiers to choose their wood and a more accurate vision of the parameters that are actually relevant for conducting the physical-mechanical and optical characterisation of the resonance woods. The combination of sensory perception and survey study permitted us to better understand the mechanism of the instrument makers' material selection. The contrast between sensory perception and opinion led us to address a variety of subjects such as the question of the intellectualisation of a gesture or some disparity between the makers' opinion and their actual action. Finally, the opinion and perception of the makers and the scientifically determined characteristics of the materials could also emphasize each other. This particular aspect is a crucial illustration of the benefit, for the future, of a decompartmentalized approach on studying the aspect of wood selection adapted to a given usage.

## Acknowledgement

This work was supported by CNRS (PhD Grant awarded to C. Carlier) and by Région Languedoc-Roussillon (Young Researcher project/prize awarded to I. Brémaud). We are grateful to the BioWooEB Research Unit of CIRAD, Montpellier, for providing access to some of their experimental facilities. Some specific characterisation could be conducted in Japan thanks to E. Obataya (Tsukuba University) and the support of JSPS, and in Italy thanks to J. and A. Sandak (CNR IVALSA) and the support of Trees4Future. Our deepest gratitude goes to violin makers who contributed to this study, for sharing their experience and ideas on the topic, and for the time they granted us in answering the survey and/or participating to the sensory tests.